# Precession Optomechanics

Xingyu Zhang*, Matthew Tomes, Tal Carmon

*Department of Electrical Engineering and Computer Science, University of Michigan, Ann Arbor, Michigan 48109, USA*
*xingyzha@umich.edu

We propose a light-structure interaction that utilizes circularly polarized light to deform a slightly bent waveguide. The mechanism allows for flipping the direction of deformation upon changing the binary polarization state of light from $-\hbar$ to $+\hbar$.

Radiation pressure has been found to cause mechanical displacement. This fact was used for controlling resonators [1–5] and waveguides [6–25]. Such forces by light in optomechanical structures were reported in the past to originate from (i) scattering forces, such as the centrifugal radiation pressure that light applies while circulating in a ring [1–3], (ii) gradient forces in resonators and waveguides [6–25], and (iii) electrostrictive pressure to excite vibration at high rates [4,5].

In waveguide structures [6–25], optical forces were suggested for all-optical reconfiguration of integrated optical devices [4], for manipulating the position of integrated optical components [6], for artificial materials [6,7] in which the internal mechanical configuration and resultant optical properties are coupled to incoming light signals [7], for tunable devices [7,8] such as actuators and transducers [8], and for more [9–25]. In all of these studies [6–25] the gradient forces are independent by the polarization state of light.

Here we suggest controlling the position of a bent waveguide with the angular momentum that light applies via its circular polarization. Unlike previous studies [6–25], no other dielectrics are needed near the waveguide and changing the polarization will flip the deformation direction.

While the linear momentum of a photon, $h/\lambda$, associated with gradient or scattering optical forces [1–25] is always along its propagation direction, a photon can also carry a different type of momentum: angular momentum [26] for which a binary vector [27] can be either with- or against- propagation. This corresponds to $+\hbar$ angular momentum for a right-handed circularly polarized [RCP] photon and $-\hbar$ angular momentum for a left-handed circularly polarized [LCP] photon. This type of angular momentum is called the intrinsic spin and is related to the fact that the photon is a spin-1 boson. We will refer to the photon in what follows as "spinning" to describe it containing angular momentum.

In the past, radiation torque effects originating from optical angular momentum were studied in waveplates [26], optically active medium [28], anisotropic crystals [29] and tweezed particles [30,31]. In more complex geometries, a spin Hall effect [32] associated with the Berry phase [33] was reported when circularly-polarized light was helically propagating in a cylinder while free to choose another trajectory in this cylinder. In this experiment, spin-dependent deflection [34] was directly observed. A natural question to ask is what the expected forces are if light is prevented from deflecting from its original trajectory, e.g. by being guided in a helical fiber. Roughly speaking, we can say that if light is bounded in its propagation in a bent fiber, then the precessing photon would "want" to deflect from its original trajectory as it described in the above experiment [34]. Yet, held by the bent fiber, light would tend to apply a force to deform this waveguide that is preventing it from changing its trajectory. In the following section, this force will be calculated from the angular-momentum conservation consideration. This type of force is general and is taken into account for the non-optical case, such as the dynamics of precessing rotors as treated in many basic textbooks [35].

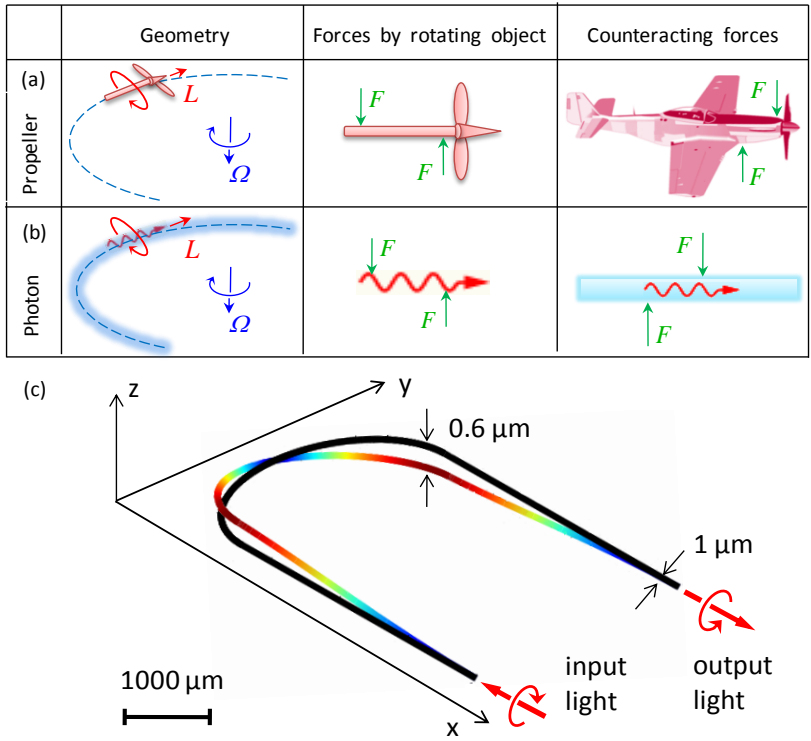

Fig. 1. Concept: (a) While precessing, changes in the orientation of the rotating shaft of a propeller airplane result in torque on what holds the shaft. (b) A spinning photon taking an identical path will apply similar torque on what holds it. (c) Illustration of the waveguide deformation due to the precession torque applied by photons. Airplane illustrated behavior is following reference [35]. The waveguide diameter is 1 µm, the bent radius is 1000 µm and the straight part is 3500 µm.

Here we propose a device in which torque, applied by spinning photons that are guided around a horizontal bend, causes mechanical deformation along the vertical direction. We name this effect "precession optomechanics", since a photon in a waveguide and a precessing rotor will apply similar torque on what holds them when taking a similar turn.

*Concepts and principles.* As shown in Fig. 1 and 2, when a propeller airplane containing angular momentum, $L$, is turning at an angular velocity, $\Omega$, then after a short time, $\Delta t$, the rotation axis has turned to a new position, tilted at an angle $\Delta\theta$. As explained in many textbooks [36], the angular momentum of the object does not change its magnitude, $L$, but it does change its direction by the amount $\Delta\theta$. The magnitude of the vector $\Delta L$ is thus $\Delta L = L\Delta\theta$ so that the torque, which is the time rate of change in angular momentum, is $\tau = \Delta L/\Delta t = L\Delta\theta/\Delta t = L\Omega$. Taking the directions of the various quantities into account, the torque that the object applies on what holds it is

$$\tau = \frac{d\bm{L}}{dt} = \bm{\Omega} \times \bm{L}. \tag{1}$$

In the example shown in Fig. 1, this torque, $\boldsymbol{\tau} = \tau \hat{\boldsymbol{r}}$ (in cylindrical coordinates), originates from an angular momentum, $\boldsymbol{L} = L\hat{\boldsymbol{\theta}}$, carried by an airplane shaft while turning at $\boldsymbol{\Omega} = \Omega\hat{\boldsymbol{z}}$. Such torque can cause an airplane to pitch, unless the pilot adjusts the control surfaces and makes use of the atmosphere to apply a counteracting force. Here, we propose to replace the propeller shaft with spinning photons that will result in a similar torque to deform a U-shaped single-mode waveguide. Such resulting waveguide deformation is illustrated in Fig. 1(c).

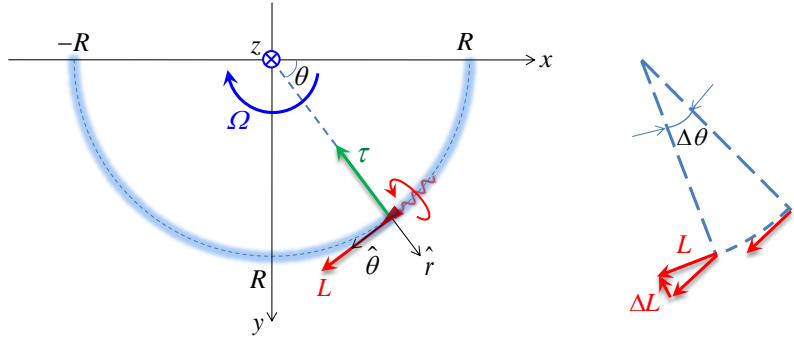

Fig. 2. Precession torque by light. Object with angular momentum $L\hat{\boldsymbol{\theta}}$ turning at angular velocity $\Omega\hat{\boldsymbol{z}}$ will apply torque $-\tau\hat{\boldsymbol{r}}$, related to the time rate of change of angular momentum, $-dL\hat{\boldsymbol{r}}/dt$.

*Calculating precession optical forces.* Optical power, $P$, entering a fiber will provide $N = \dfrac{P}{\omega\hbar}$ photons per second, where $\omega$ is the angular frequency of light. Taking $c/n_{eff}$ for the speed of light in the fiber, this waveguide will contain $\dfrac{dN}{dx} = \dfrac{P}{\omega\hbar}\dfrac{n_{eff}}{c}$ photons per meter. $n_{eff}$ is the effective refractive index of the waveguide, taking into account the core index, the cladding index, and the core diameter as described in [37]. Considering that each photon carries $\hbar\hat{\boldsymbol{\theta}}$ angular momentum, the intra-fiber angular momentum per unit length becomes $\dfrac{d\boldsymbol{L}}{dx} = \dfrac{P}{\omega\hbar}\dfrac{n_{eff}}{c}\hbar\hat{\boldsymbol{\theta}}$. As long as these photons are propagating in a straight fiber, their turning angular velocity $\boldsymbol{\Omega}$ is zero and they exert no precession forces. Yet, when they enter a curve of radius, $r$, their angular velocity becomes $\boldsymbol{\Omega} = \dfrac{c}{n_{eff}r}\hat{\boldsymbol{z}}$. We can then use Eq. (1) and get the precession torque per unit length, $dx$, as

$$\frac{d\boldsymbol{\tau}}{dx} = \boldsymbol{\Omega} \times \frac{d\boldsymbol{L}}{dx} = \frac{c}{n_{eff}r}\hat{\boldsymbol{z}} \times \frac{P}{\omega\hbar}\frac{n_{eff}}{c}(\pm\hbar)\hat{\boldsymbol{\theta}} = \mp\frac{P}{r\omega}\hat{\boldsymbol{r}} = \mp\frac{P}{c}\frac{\lambda}{2\pi r}\hat{\boldsymbol{r}}. \quad (2)$$

We have added here the sign $\pm$ to describe either right-handed or left-handed circularly polarized light, $\lambda$ denotes the vacuum wavelength, and $c$ is the vacuum speed of light. For a bend as in Fig. 1(c) with a 1 Watt optical input at 1.55 micron wavelength the distributed torque will be in the order of $10^{-12}\ \text{N}\cdot\text{m}/\text{m}$. Interestingly, while the linear momentum of photon scales inversely with wavelength, its angular momentum is wavelength independent. It is better hence to use long-wavelength light, so that each Watt of optical power will carry more photons and consequently greater total angular momentum. Accordingly, the torque in Eq. (2) is indeed scaled linearly with the wavelength. Experimentally, it means that IR telecom-compatible light will produce three times the deformation in comparison with visible light. We choose here to calculate the distributed torque, $d\tau/dx$ (Eq. (2), units of Newton·meter/meter), in order to substitute it in as the external load applied on a bent waveguide and check for the consequent deformation.

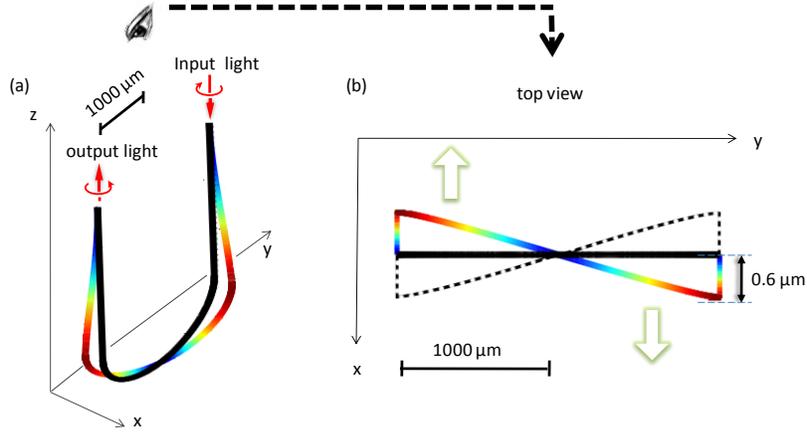

Fig. 3. Proposed experiment. (a) 3-D perspective of circularly polarized light entering one end of a U-shaped waveguide and exiting from the other end, causing the waveguide to deflect from its mechanical equilibrium. The colored curve indicates the deflected shape and the colors stand for deformation amplitude. The black curve indicates the original U shape at mechanical equilibrium. The amplitudes of deformation are exaggerated for clarity. (b) Top-view of the deformation as expected to be seen by a standard microscope. The dashed curve indicates the deflected shape caused by switching polarization from RCP light to LCP light. Videos for binary deformation flipping in 3-D perspective and top view are shown in Media 1 and Media 2 separately. Device dimensions are as in Fig. 1(c).

*Calculating deformation.* In what follows we will calculate deformation for the longest and thinnest tapered fiber that we can experimentally fabricate while maintaining transmission higher than 90%. The ratio between the bent and the straight part will be then optimized for maximal deformation. To calculate the deformation of the structure resulting from the precession forces by light, we solve the tensorial strain-stress equation using a finite element method [38]. The analyzed structure is a single-mode taper [39,40], bent similar to that in reference [41], but with dimensions shown in Fig. 1(c). The boundary conditions applied on this structure are distributed torque (Eq. (2) for the bent regions, zero load at the straight regions and zero motion at the ends where they are mechanically mounted. Upon deformation, the torque from Eq. (2) is modified as a function of the local deformed radius. We iteratively take this torque modification (due to the deformation) into account and stop iterating when the deformation accuracy is better that 0.1%. We comment that fabrication and measurement tolerances in such experiments are typically larger than 1%. Energy loss due to the loop radius was calculated for our case [42] to be less than 1/million. Such low loss is typical to bent structure [43] where the index contrast between the silica core and the air cladding is large.

The major result is shown in Fig. 3 demonstrating that an input optical power of 1 Watt at IR wavelength, $\lambda = 1.55\,\mu m$, causes micron-scaled deformation. A standard erbium amplifier can provide such a 1 Watt output, and a simple microscope can monitor the micron-scaled motion. One possible experiment is to look at the region where deformation is maximal (red color region in Fig. 3) while changing the polarization from clockwise to counter clockwise, this region of the fiber is expected to move forth and back. Being careful, we should also verify that the circular polarization does not change as propagating along the bend. In principal, the polarization is expected to change because the bent waveguide boundary conditions are discriminating between the horizontal polarization and the vertical polarization states. These modes are hence delayed in respect with each other to modify polarization. As these modes are superimposed to produce the circularly polarized light, the circular polarization is expected to be changed to some extent. We hence numerically calculate [42] the propagation constants for the vertical- and horizontal-polarization modes ($E_r \hat{r}$ and $E_z \hat{z}$) and got effective refractive indices that are different by $10^{-8}$. This implies that there is some structural birefringence. Equating for propagation along the bend as in Fig. 1(c), a circular polarization $E_r \hat{r} + i \cdot E_z \hat{z}$ will be slightly modified to $E_r \hat{r} + (0.0001 + 0.99999\,i) \cdot E_z \hat{z}$ at the end of the bend, implying that the loss of circular polarization here is negligible.

*Binary deformation flipping.* Not being limited in the applied torque direction, switching the polarization from RCP to LCP is calculated to allow for binary flipping of the deformation direction, as shown by the dashed line in Fig. 3(b).

*Optimization of deflection.* A major goal in this calculation is preparing the ground for an experiment to be performed. One task here is, hence, to optimize deformation to be maximal. Typical to such structures, the thinner and longer the structure, the larger the deformation. We hence choose to make the waveguide for this calculation thin (1 μm) and long (10mm), similar to what can be fabricated with current technology [39,40,44]. With this given waveguide we choose "U" shape that from our experience is easy to create. A proper combination of bent- and straight-elements is needed to allow the maximal deformation. This is because the bent structure provides the torque (that is proportional to $1/r$) while the straight section provides the softness. For a typical experimental example, a 10mm waveguide bent into a U shape having a 0.75 mm radius (for the U semicircle) will provide the maximal deformation as we calculated numerically (Fig. 4).

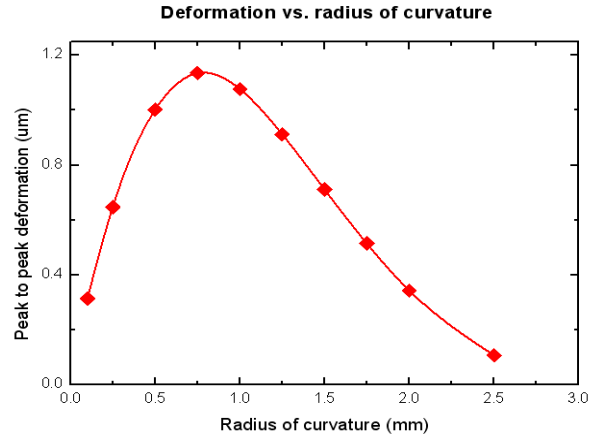

Fig. 4. Optimization of the deformation. When the radius of curvature is around 0.75mm, the peak to peak amplitude of the deformation is greater than 1 μm. The input optical power is 1Watt at 1.55 μm, the diameter of the tapered waveguide is 1 μm and the total length is 10mm.

*Competition with other effects.* It is important to prove that precession optical forces are dominating in our structure and are not disturbed by other effects. To show that, we repeat our calculation from Fig. 3, but this time we also take into account centrifugal radiation pressure [1–3]. The centrifugal radiation pressure is given by [1]

$$\frac{d\mathbf{F}}{dx} = \frac{2Pn_{eff}}{c\pi r}\hat{\mathbf{r}}. \qquad (3)$$

In the proposed experiment, the input optical power, $P$, is 1 Watt, the effective refractive index, $n_{eff}$, is 1.2 [42], and the radius of curvature, $r$, is 1mm. Based on Eq. (3), the centrifugal radiation pressure per unit length is calculated to be $2.546 \times 10^{-6}\,\mathrm{N\cdot m/m}$. The direction of centrifugal radiation pressure is along the radius, pointing out of the center. To mimic a typical experiment, we also add gravity. Figure 5 shows that the relevant precession deformation (in the direction of the arrows) dominates by being more than 3 orders of magnitude larger than the strain caused by centrifugal radiation pressure and gravity combined. Further, centrifugal radiation pressure is always radially outward, irrespective of the polarization. On the contrary, precession effects are polarization-dependent and the deformation will flip upon switching the binary polarization state of light from RCP to LCP (Fig. 3).

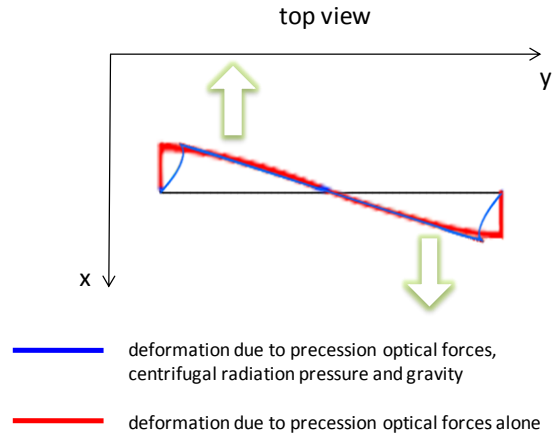

Fig. 5. Verifying no disturbing effects by centrifugal radiation pressure and gravity. Precession deformation is nearly unaffected by centrifugal radiation pressure or gravity. The red curve indicates the precession deformation alone, and the blue curve indicates deformation with the addition of centrifugal radiation pressure and gravity. Device dimensions are as in Fig. 1(c).

As for thermal effects, unlike the optomechanical precession, they are expected to be polarization independent and hence not to flip upon polarization switching. Further, the thermal expansion is expected to be orthogonal with the precession deformation calculated in Fig. 3. It will be easy to cancel out thermal effects in our experiment by simply holding the input power constant and just changing the polarization state of input light from right- to left-handed circular polarization, in this way the temperature and thermal deformation are not expected to change, while the deformation we predict here is expected to switch.

*Energy consideration.* In the transition period from mechanical equilibrium to a twisted shape, precession optical forces are applied against a moving surface and energy is taken from light to load the mechanical spring. As no photons are absorbed in this process, a fundamental question arises concerning the mechanism that converts energy from electromagnetic to mechanical. In such a situation, circularly polarized light is experiencing Doppler shift caused by rotation. This type of energy conversion is called "rotational frequency shift" [45,46], in contrast with the common Doppler shift that is typically caused by reflection from a moving, non-rotating body. Like a suspended mirror experiencing radiation pressure [47], no scattering, de-coherence, or rotational Doppler shift are expected after the system reaches a steady state and the deformation is balanced by precession optical forces.

*Conclusion.* Here we calculated that a set of two opposite device deformations is excited by switching the binary spin-state of photons. This effect can be considerably amplified, for example, by switching the polarization state of light from RCP to LCP at the eigen mechanical frequency (~200Hz) of the bent waveguide as in Fig. 1(c). The higher the mechanical quality factor of this oscillator, the greater the deformation.

Additionally, putting optical reflectors on both sides of the bent tapered fiber to turn it into a Fabry-Perot resonator [48] will enhance deformation to scale with the finesse of this optical cavity. Interestingly, the deformation is enhanced through photon reflections from the Fabry-Perot ends, as both angular velocity, $\Omega$, and spin, $L$, change their signs so that the torque product, $\tau = \Omega \times L$, does not change its sign and is resonantly enhanced.

In a more general treatment, the angular momentum of light can be amplified by utilizing the orbital angular momentum of Laguerre-Gaussian modes where the angular momentum is associated with helical wavefronts of the propagating light [49,50]. Such beams were demonstrated with angular momentums greater than $150\,\hbar$ per photon [51,52] and suggest corresponding enhancement of the precession optomechanical deformation. The tradeoff is that a different waveguide will be needed for such Laguerre-Gaussian modes.

Except for having a binary character and originating from angular momentum rather than linear momentum, the precession optical force proposed here is different from other ways to control the position of optical waveguides, as it does not need another waveguide [6–8,10–24] or bulk material [9,25] next to it and can be completely isolated from other dielectrics.

## Acknowledgments

This work was supported by the DARPA ORCHID program through a grant from AFOSR.